\documentclass[fleqn,10pt]{wlscirep}
\usepackage[utf8]{inputenc}
\usepackage[T1]{fontenc}
\usepackage{xcolor}
\newcommand{\ER}{Erd\H{o}s-R\'enyi}
\DeclareUnicodeCharacter{2212}{\ensuremath{-}}

\title{Effective Noise Thresholds for Reliable Quantum Approximation Optimisation Algorithm under Noisy Environment}
\author[1,2, *]{Yozef Tjandra}
\author[2]{Hendrik Santoso Sugiarto}
\author[2]{Celine Vania Setiadi}
\author[2]{Joanna Gracia Tan}
\author[3]{Andriyan Bayu Suksmono}
\affil[1]{Doctoral Program of Electrical Engineering and Informatics, School of Electrical Engineering and Informatics (STEI), Institut Teknologi Bandung, Bandung, 40132 Indonesia.}
\affil[2]{Department of IT \& Big Data Analytics (IBDA), Calvin Institute of Technology, Jakarta, 10610, Indonesia}
\affil[3]{The School of Electrical Engineering and Informatics (STEI), Institut Teknologi Bandung, Bandung, 40132 Indonesia.}

\affil[*]{Corresponding author: Yozef Tjandra, email: yozef.tjandra@calvin.ac.id}


\begin{abstract}
Variational quantum algorithms are leading candidates for practical utilisation on noisy intermediate-scale quantum (NISQ) processors, but accumulating noise bounds the usefulness of circuit depths. For the quantum approximate optimisation algorithm (QAOA), this trade-off has mostly been addressed by searching for the best depth for a given device. In this work, we consider the inverse question: how much noise each depth can tolerate. Using GPU-accelerated noisy QAOA simulations for MaxCut on three graph families with 6 to 18 vertices, depths up to 10 and three noise channels, we fit the approximation ratios with a generalised logistic surface. We propose the \textit{effective noise threshold} as the inflection boundary of the logistic surfaces, a noise boundary separating reliable and unreliable QAOA approximation regimes. In general, the threshold decreases as depth rises, indicating that deeper circuits demand for stricter noise requirement for reliable approximation. Its best and worst values over depth divide the noise axis into three regions: every depth remains effective, depth must be chosen with care, or no depth induces effective optimisation.Assessing on the \texttt{ibm\_marrakesh} processor, although modelled gate errors understate factual noise, the threshold modeling remains beneficial in informing qualitative depth-dependence structure and identifying dominating noise channel in noisy QAOA.
\end{abstract}

\begin{document}

\flushbottom
\maketitle
%
%
\thispagestyle{empty}


\section*{Introduction}

Combinatorial optimisation has emerged as one of the most promising domain for demonstrating practical quantum advantage on noisy intermediate-scale quantum (NISQ) devices~\cite{abbas2024challenges,pirnay2024principle}. Many industrial problems can be formulated as NP-hard combinatorial optimisation tasks whose exact classical solution is generally believed to scale exponentially with the instance size. Consequently, many works aim to design various quantum algorithms exploiting superposition and entanglement phenomena in quantum mechanics to generate high-quality approximate solutions, which had been succesfully demonstrated in various areas including logistics\cite{moncayo2025quantum}, finance\cite{yuan2026quantifying}, and energy systems\cite{yeo2025quantum}.

Quantum optimisation scheme has developed from adiabatic quantum computation (AQC) towards the more universal gate-based variational approaches. In AQC, the optimisation process leverages a quantum system's slow evolution towards the ground state of a problem Hamiltonian~\cite{albash2018adiabatic}. Whereas, in Variational Quantum Algorithm (VQA), a parameterised circuit is tuned iteratively by a classical optimiser~\cite{cerezo2021variational} to perform various tasks, including combinatorial optimisation. Their comparatively shallow circuits and the use of classical feedback loops make VQA attractive candidates for resource-limited NISQ hardware.

\begin{figure}[ht]
    \centering
    \includegraphics[width=0.8\linewidth]{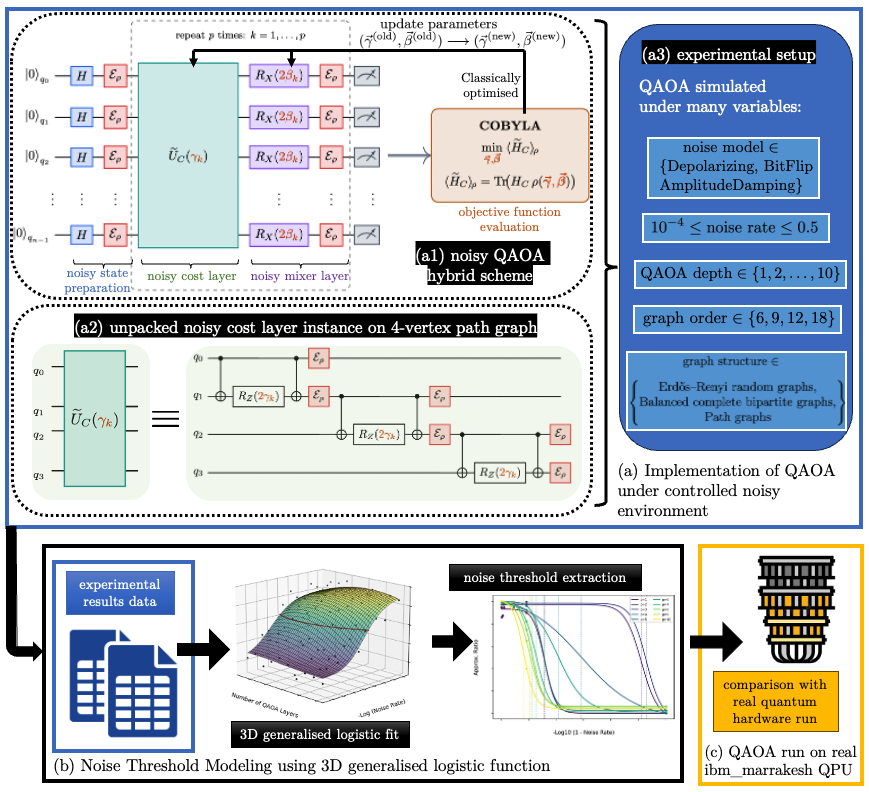}
    \caption{Overview of the noisy QAOA workflow and analysis pipeline used in this study. \textbf{(a)} Implementation overview of simulated QAOA under controlled noise. \textbf{(a1)} The generic scheme of the hybrid quantum-classical noisy QAOA: noisy state preparation is followed by $p$ repetitions of the noisy cost and mixer layers, with the noise channel $\mathcal{E}_\rho$ applied after each gate; measurement outcomes are used to evaluate the objective function $\langle \tilde{H}_C\rangle_{\rho}$ which is minimised over the variational parameters $(\vec{\gamma}, \vec{\beta})$ using the COBYLA classical optimiser. \textbf{(a2)} Example decomposition of the noisy cost layer $\tilde{U}_C(\gamma_k)$ into its constituent two-qubit entangling gates and associated noise channels, illustrated for a four-vertex path graph. \textbf{(a3)} Summary of the experimental variables spanned in the controlled simulation study. \textbf{(b)} Noise threshold modeling: the QAOA solution approximation ratio data resulted from the noisy simulation experiment are fitted into a three-dimensional generalised logistic surface as a function of QAOA depth and $-\log(\text{noise rate})$. \textbf{(c)} The extracted thresholds are subsequently assessed under the real \texttt{ibm\_marrakesh} superconducting quantum processor.}
    \label{fig:noisy-qaoa}
\end{figure}

The Quantum Approximate Optimisation Algorithm (QAOA)~\cite{farhi2014quantum} is one variational framework designed specifically for this setting. QAOA digitises adiabatic evolution into $p$ layers of alternating unitaries, namely the problem and mixing unitaries (see Figure~\ref{fig:noisy-qaoa}.a.1). For each unitary block, real-valued parameters are binded within them and their values are updated in a loop using a classical optimiser. In a noiseless environment, the best achievable solution cannot worsen as $p$ increases, so depth is the key operational resource to improve the quality of the optimisation solution. However, each additional layer adds more entangling gates in proportion to the size of the problem instance (see Figure~\ref{fig:noisy-qaoa}.a.2), so that on real physical hardware, significant cumulated errors are expected as $p$ grows. Greater circuit expressivity therefore competes with noise-induced degradation, implying that the choice of the QAOA depth $p$ heavily dependent on the device noise level.

Many existing works confirm that the benefit of adding more QAOA layers is not guaranteed on noisy devices. Several noisy QAOA simulation on various combinatorial problems such as minimum dominating set~\cite{li2026auxiliary}, vertex cover~\cite{pellow2024effect}, and MaxCut~\cite{liu2025noisy} have reported numerical experimental results indicating that deeper QAOA circuits often experience diminishing returns.  Experiments on real quantum processors also point in the same direction, with noise limiting the gain from added layers~\cite{harrigan2021quantum,weidenfeller2022scaling,suksmono2025quantum,ebadi2022quantum}. Furthermore, it is also known that the combinatorial problem structure itself also affects the degree of severity where noise degrades the optimisation solution\cite{thelen2024approximating}. These studies establish that an optimal depth exists, but each characterises it for a limited set of noise settings, problem instances or devices.

Several theoretical studies offer a complementary explanation. Local noise drives the output of a noisy QAOA circuit towards that of the maximally mixed state as the number of noisy gates grows~\cite{marshall2020characterizing,franca2021limitations}. Moreover, the application of local Pauli noise in parameterised quantum circuits induces parameter gradients that vanish exponentially with circuit depth during the classical optimisation iterations~\cite{wang2021noise}, truncating only the last logarithmic layers that remain effectively trainable~\cite{mele2026noise}. These results suggests that noise imposes an effective QAOA depth budget whose value shrinks as the error rate rises. However, most of these results are asymptotic statements which do not indicate at which finite noise rate a given depth ceases to be useful for a concrete problem instance.

Most responses to this tension treat the noise of the device as a given constant and inquire for which QAOA depth is the most optimal to use. Parameter-transfer heuristics that initialise depth $p+1$ from optimised depth-$p$ solutions~\cite{zhou2020quantum} lower the cost of scanning successive QAOA depths. Adaptive construction strategy also exists: growing the ansatz iteratively using problem-tailored elements~\cite{zhu2022adaptive}; and automatically selecting the optimal QAOA depth through $l_1$-regularisation under noise~\cite{pan2022efficient} or adiabatic interpolation~\cite{saini2025dynamic}. Layerwise learning mitigates barren plateaus by training a circuit block by block, adding layers only after the shallower circuit has been optimised~\cite{skolik2021layerwise}, whereas Quantum Architecture Search (QAS) treats gate composition and connectivity as quantities to be optimised, in some formulations under hardware noise~\cite{zhang2022differentiable,du2022quantum}. All of these approaches address the forward problem of which depth or ansatz performs most optimally on a given device. To the best of our knowledge, no work has identified how much noise rate that a given QAOA depth can tolerate to maintain effective optimisation.

Systematic studies of noisy QAOA implementation have varied some essential factors that successfully generate meaningful findings, but not in a form that that shows at which noise rate a given QAOA depth ceases to be useful. Several examine performance at fixed noise strengths~\cite{pellow2024effect,lotshaw2021empirical}, pointing out that an optimal depth exists but not how it moves as the noise changes. Whereas, others systematically assess various noise rates but only on a limited set of problem instances and QAOA depths~\cite{xue2021effects}. Although some experimental works on real quantum devices exist\cite{irfan2025quantum}, they remain limited due to expensive quantum resources and calibration instability, so the relation between simulated noise and device behaviour remains largely untested. What is missing is the inverse of the depth-selection problem: rather than asking which depth suits a given device, it is insightful to ask how much noise a given depth can tolerate, and how this tolerance moves with QAOA depth, noise channel and instances. 

In this work, we introduce the notion of \textit{effective noise threshold} $r^\star(p)$, the noise level at which the approximation ratio of the optimisation solution from a depth-$p$ QAOA circuit transitions from a high-performance to a noise-dominated regime. Such a quantity would serve both as a rule for both choosing depth before execution and as a target for hardware error rates. We define it operationally as the inflection boundary of a generalised logistic surface fitted to the approximation ratio as a joint function of the QAOA depth and noise rate. This boundary curve marks the situation of greatest sensitivity to noise. 

Using GPU-accelerated noisy simulation of MaxCut on Erd\H{o}s--R\'enyi, path, and complete bipartite graphs with up to 18 vertices, we apply depolarising, bit-flip, and amplitude damping noise channels to the QAOA circuits, and extract $r^\star(p)$ for depths up to $p=10$. We find that the threshold generally falls as depth increases whose value depends on the noise channel, graph size and density. Moreover a noise budget for a target depth can therefore be read directly from the fitted surface. We then assess the utility potential of the extracted thresholds using executions under the \texttt{ibm\_marrakesh} superconducting processor. The depth dependence of the hardware results follows the structure of the thresholds, while the overall performance level indicates that the modeled gate error understates the factual noise of the hardware circuit. 

In summary, our work provides a complementary view of the QAOA depth analysis: it reframes the question about how to choose the optimal QAOA depth into a question about the noise budget of a device for a given set of quantum resource, with the threshold $r^\star(p)$ giving a quantitative estimate of that budget. Figure \ref{fig:noisy-qaoa} provides the overall pipeline of this work.

\section*{Preliminaries: QAOA Foundations and Noise Sensitivity}

\subsection*{The MaxCut Problem and QAOA Formulation}

The Maximum Cut (MaxCut) problem is a canonical NP-hard combinatorial optimisation problem and serves as the primary benchmark throughout this work. Given an unweighted, connected, and undirected graph $G = (V, E)$, the problem's objective is to partition $V$ into two disjoint subsets that maximises the number of cut edges whose end vertices sit on different subsets. Assigning each vertex $u\in V$ a spin variable $z_u\in\{-1,+1\}$, the number of cut edges is \begin{equation} C(\textbf{z})=\sum_{(u,v)\in E}\frac{1-z_u z_v}{2}, \label{eq:maxcut} \end{equation} and its maximum value is denoted by $C_{\mathrm{opt}}=\max\limits_{\textbf{z}} C(\textbf{z})$.

Replacing each $z_u$ by the Pauli-$Z$ operator on qubit $u$ gives the diagonal problem Hamiltonian \begin{equation} H_C=\sum_{(u,v)\in E}\frac{-(1-Z_u Z_v)}{2}, \label{eq:hc} \end{equation} which satisfies $H_C|\textbf{z}\rangle=-C(\textbf{z})|\textbf{z}\rangle$. The sign is chosen so that the maximum cut is the ground state of $H_C$, and MaxCut becomes the minimisation of $\langle H_C\rangle$.

QAOA~\cite{farhi2014quantum} prepares a state of this kind with $p$ alternating layers of the problem unitary $U_C(\gamma)=e^{-i\gamma H_C}$ and the mixing unitary $U_M(\beta)=e^{-i\beta H_M}$, where $H_M=\sum_{u\in V}X_u$ is the standard Pauli-$X$ mixer: \begin{equation} |\psi(\vec\gamma,\vec\beta)\rangle = U_M(\beta_p)\,U_C(\gamma_p)\cdots U_M(\beta_1)\,U_C(\gamma_1)\,|+\rangle^{\otimes |V|}, \label{eq:qaoa_state} \end{equation} with $\vec\gamma,\vec\beta\in\mathbb{R}^p$. A classical optimiser tunes these $2p$ real-valued parameters to minimise $\langle H_C\rangle=\langle\psi(\vec\gamma,\vec\beta)|H_C|\psi(\vec\gamma,\vec\beta)\rangle$. The optimised circuit is then sampled to yield a solution candidate. The solution quality is evaluated through the approximation ratio metrics defined in the Experimental Setup, which is based on the most frequently sampled bitstring rather than on $\langle H_C\rangle$ itself.

Two properties of the depth $p$ are central to this work. First, the optimum of $\langle H_C\rangle$ cannot increase with $p$ in the noiseless case, because setting $\gamma_p=\beta_p=0$ recovers the depth-$(p-1)$ circuit. In the limit $p\to\infty$ with optimal parameters, QAOA recovers $\langle H_C\rangle=-C_{\mathrm{opt}}$, as the layered structure approximates a Trotterised adiabatic evolution~\cite{farhi2014quantum}. Second, in the standard decomposition each $U_C(\gamma)$ requires two CNOT gates per edge, so a depth-$p$ circuit contains $2|E|p$ entangling gates. Circuit noise exposure thus grows with both the depth and the density of the graph, which underlies the competition studied here.

\subsection*{Quantum Noise Models}

In realistic quantum hardware, quantum states are inevitably subject to environmental interactions that might corrupt the ideal unitary evolution assumed in theoretical analyses. In this work, we employ the widely known noise models represented by completely positive and trace-preserving maps on the density matrix $\rho$, written in Kraus form as $\mathcal{E}(\rho)=\sum_k E_k\rho E_k^\dagger$, where the operators $\{E_k\}$ satisfy the trace-preservation condition $\sum_k E_k^\dagger E_k=I$. We consider the depolarising, bitflip, and amplitude damping noise channels, each controlled by a common noise rate $r\in[0,1]$, described respectively as follows: \begin{align} \mathcal{E}_{\text{D}}(\rho) &= (1-r)\rho + \frac{r}{3}\left(X\rho X + Y\rho Y + Z\rho Z\right), \label{eq:depolarising} \\ \mathcal{E}_{\text{BF}}(\rho) &= (1-r)\rho + r X\rho X, \label{eq:bitflip} \\ \mathcal{E}_{\text{AD}}(\rho) &= E_0\rho E_0^\dagger + E_1\rho E_1^\dagger, \quad E_0 = \begin{pmatrix} 1 & 0 \\ 0 & \sqrt{1-r} \end{pmatrix}, \quad E_1 = \begin{pmatrix} 0 & \sqrt{r} \\ 0 & 0 \end{pmatrix}. \label{eq:amplitude} \end{align} The depolarising channel applies each of the Pauli errors $X$, $Y$ and $Z$ with probability $r/3$, and the bit-flip channel applies $X$ with probability $r$. Both are unital, so they drive the state towards the maximally mixed state as errors accumulate~\cite{marshall2020characterizing,franca2021limitations}. The amplitude damping channel models energy relaxation, in which $|1\rangle$ decays to $|0\rangle$ with probability $r$. It is non-unital and drives the register towards $|0\rangle^{\otimes n}$, which places all vertices in the same partition and therefore has a cut value of zero. These channels thus represent two distinct error mechanisms: unbiased stochastic Pauli errors and biased relaxation. 

In the noisy simulations, the selected channel, denoted generically by $\mathcal{E}_\rho$, is applied with the same rate $r$ after every gate, including the state preparation layer, on each qubit that the gate acts on. A two-qubit gate is therefore followed by an independent single-qubit channel on each of its two qubits. Both one-qubit and two-qubit gates share the same $r$. Errors are assumed to be Markovian and uncorrelated across qubits and gates.

Let $\tilde U_C(\gamma)$ and $\tilde U_M(\beta)$ denote the compositions of the noisy gates in the decompositions of $U_C(\gamma)$ and $U_M(\beta)$. To implement the noisy QAOA simulation, the pure state of Eq.~\eqref{eq:qaoa_state} is then replaced by the density matrix 
\begin{equation} \rho(\vec\gamma,\vec\beta)=\tilde U_M(\beta_p)\circ\tilde U_C(\gamma_p)\circ\cdots\circ\tilde U_M(\beta_1)\circ\tilde U_C(\gamma_1)\left(\tilde\rho_0\right), \label{eq:noisy_state} \end{equation}
where $\tilde\rho_0$ is the noisy preparation of $|+\rangle^{\otimes |V|}$. On the noisy setup, the optimised objective becomes $\langle\tilde H_C\rangle_\rho=\mathrm{Tr}\left(H_C\,\rho(\vec\gamma,\vec\beta)\right)$.

\subsection*{Critical Depth and Effective Noise Threshold for Noisy QAOA}
Under noiseless setup the minimum of $\langle H_C\rangle$ cannot increase as $p$ grows, however this guarantee does not extend to noisy execution. Each additional layer adds gates, so the accumulated error grows with $p$. Under depolarising noise, the output of a deep circuit approaches the maximally mixed state~\cite{marshall2020characterizing,franca2021limitations}, and for local Pauli noise the gradients of the cost function vanish exponentially with circuit depth~\cite{wang2021noise}. At a fixed noise rate $r$, the approximation ratio (AR), the optimisation performance metrics (see Subsection Experimental variables and constants for the exact definition), is therefore generally non-monotonic in $p$. We define the \textit{critical depth} $p_{\mathrm{crit}}(r)$ as the depth that maximises the AR averaged over independent runs at noise rate $r$, beyond which additional layers are counterproductive.

The critical depth addresses a forward question: which depth is best for a given noise rate as investigated and exposed by many existing works\cite{pellow2024effect,pan2022efficient,saini2025dynamic}. We instead pose the inverse question: at a fixed depth $p$, how much noise can the circuit tolerate? We formalise this quantity as the \textit{effective noise threshold} $r^\star(p)$, the noise rate at which the AR of a depth-$p$ QAOA circuit passes from a high-performance regime to a noise-dominated regime.

\section*{Experimental Setup}
\subsection*{Graph instances}
\begin{figure}
    \centering
    \includegraphics[width=0.8\linewidth]{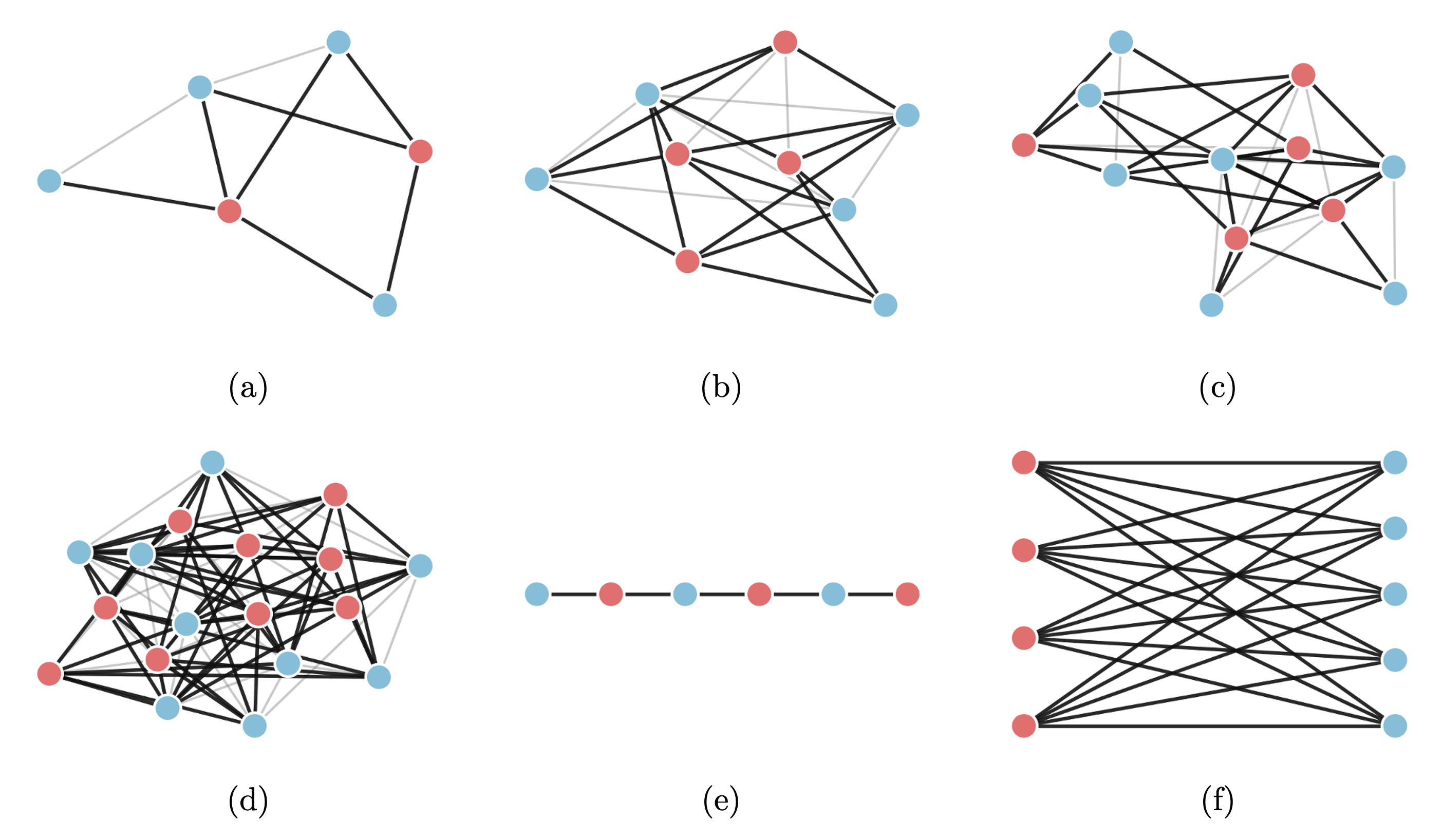}
    \caption{Representative graph instances used for QAOA benchmarking. Node colours indicate a sample of optimal MaxCut bipartition solution. Bold edges denote cut edges in the optimal solution, whilst grey edges are intra-partition edges. Instances of Erd\H{o}s--R\'{e}nyi random graphs with \textbf{(a)} 6 nodes, \textbf{(b)} 9 nodes, \textbf{(c)} 12 nodes, and \textbf{(d)} 18 nodes. \textbf{(e)} a representative of the path graph family with 6 nodes. \textbf{(f)} a representative of the complete balanced bipartite graph family $K_{4,5}$ with $9$ nodes.} 
    \label{fig:graphs}
\end{figure}

To evaluate QAOA performance across diverse structural properties, we consider three representative graph families: Erd\H{o}s--R\'{e}nyi (ER) random graphs, path graphs, and complete balanced bipartite graphs. These families are selected to capture a spectrum of edge densities and structural regularities. ER random graphs are generated using two parameters: the number of vertices and the probability of observing an edge for any possible pair of vertices within the graph\cite{erdds1959random}. In this work we set the edge probability $p_{\mathrm{ER}} = 0.6$, providing moderately dense instances with non-trivial optimisation landscapes. Path graphs represent a sparse, highly structured topology serving as a baseline for minimally connected instances. Complete balanced bipartite graphs, constructed by partitioning vertices into two subsets of equal or near-equal size with all cross-partition edges included, introduce strong symmetry known to influence variational algorithm behaviour. For each family, instances of $|V| \in \{6, 9, 12, 18\}$ vertices are generated, a range chosen to remain classically simulable while permitting systematic assessment of scaling behaviour. Representative instances are illustrated in Figure~\ref{fig:graphs}.
\subsection*{Experimental variables and constants}
\label{subseq:experimental_variables}
All controlled simulations are conducted using the NVIDIA CUDA-Q framework on various NVIDIA GPUs ranging from RTX3070,  RTX4080, and RTX4090 to allow running several simulation configurations simultaneously. Classical optimisation employs the COBYLA optimiser, and the mixing Hamiltonian is fixed to the standard Pauli-$X$ choice. Each QAOA configuration is evaluated over 10 independent runs with seeds incremented from 0 to 9, ensuring statistical reproducibility and accounting for stochastic variability in both the optimisation trajectory and measurement sampling. Performance is assessed using the approximation ratio (AR), defined as $\mathrm{AR} = C(\textbf{z}^\star) / C_{\mathrm{opt}}$, where $\textbf{z}^\star$ is the most frequently sampled solution from the final optimised circuit and $C_{\mathrm{opt}}$ is the exact optimum obtained by exhaustive classical enumeration over all $2^n$ possible solutions.

The primary experimental variables are noise rate $r\in \{0.0001, 0.0005, 0.001, 0.005, 0.01, 0.05, 0.1, 0.5\},$ and QAOA circuit depth $1\le p \leq 10$  varied jointly to characterise the trade-off between ansatz expressivity and noise accumulation. Three noise channels are considered: depolarising, amplitude damping, and bit-flip. Graph size and structure are varied as secondary variables to contextualise the generality of the findings across different problem instances. Full technical details of the QAOA implementation setup are provided in the Methods section.

\section*{Results}

\subsection*{QAOA on controlled noise environment}
We now turn to the results obtained from the implementation of QAOA with simulated noise. The complete AR landscapes for all 36 combinations of noise channel, graph family and graph order are provided as heatmaps of the run-averaged AR (Supplementary Figure 1) and of its standard deviation across runs (Supplementary Figure 2) in Section A of the Supplementary Information file. To maintain clarity and avoid main text overloading, we will discuss some representative slices of the results.

\begin{figure}[ht]
    \centering
    \includegraphics[width=0.95\linewidth]{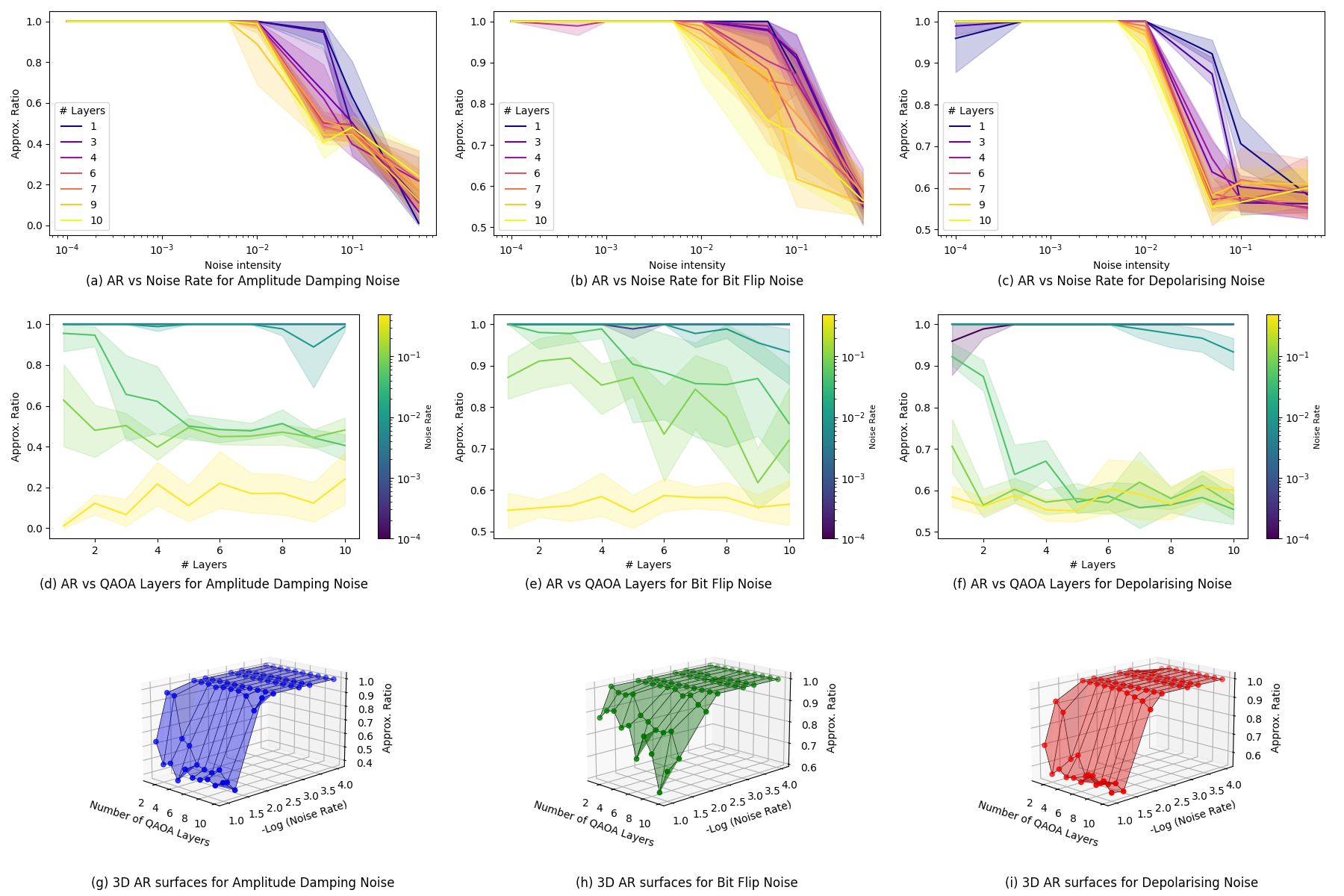}
	\caption{QAOA approximation ratio (AR) under controlled noise for the 18-vertex complete bipartite graph $K_{9,9}$, across various QAOA depths, noise rates and noise channel types. \textbf{(a--c)} AR as a function of noise intensity (log scale) for the amplitude damping, bit-flip and depolarising channels respectively; shown for various QAOA depths. \textbf{(d--f)} The corresponding AR as a function of QAOA depth, coloured by noise rate, for the same three channels. \textbf{(g--i)} 3D surfaces visualising the AR as a function of QAOA depth and $−\log(\text{noise rate})$, exposing the joint dependence appearing in \textbf{(a--f)}.}
    \label{fig:2D_noise_bipartite_18}
\end{figure}

To provide a snapshot exposing QAOA's behaviour under varying noise levels and circuit depths, we present Figure~\ref{fig:2D_noise_bipartite_18} corresponding to the instance of $K_{9,9}$, the complete bipartite graph with 18 vertices. This instance particularly serves as a representative example of trends that appear across various graph configurations. 

From the first row of subplots within the figure, it is evident that across all noise types and QAOA depths, the approximation ratio (AR) generally degrades as the noise rate increases. However, the degradation is not linearly gradual as the AR often remains relatively stable at low noise rates before undergoing a sudden decline beyond a certain region. This behaviour suggests the existence of a transition region separating high and low efficacy of QAOA.

The dependence of AR on QAOA depths exhibits qualitatively different behaviours across different noise regimes, as observed in the second row subplots of Figure \ref{fig:2D_noise_bipartite_18}. In the low-noise regime, the AR remains close to unity for shallow circuits while slightly degrade for ones with deeper depths. In contrast, under sufficiently strong noise, any choice of circuit depths yields significantly lower AR values due to accumulated decoherence and gate errors. Interestingly, in an intermediate noise regime, a non-trivial behaviour is observed: for a subset of noise rates, a local AR maximum is identifiable at a certain number of layers. This indicates the presence of an optimal circuit depth, beyond which further increases in depth become detrimental. Such behaviour highlights the importance of carefully balancing circuit depth and noise, rather than naively increasing $p$, in order to achieve optimal QAOA performance in realistic noisy settings.

While comparing the columns of Figure \ref{fig:2D_noise_bipartite_18}, a notable distinction emerges when comparing the three noise models. Amplitude damping induces the most severe deterioration, eventually driving the AR close to zero at sufficiently high noise rates. In contrast, depolarising and bit-flip channels tend to saturate at non-zero AR values even under very strong noise ($r=0.5$). This distinction reflects the dissipative nature of amplitude damping, which continuously relaxes qubits toward the ground state and progressively erases problem-encoded quantum information.

The corresponding three-dimensional AR surfaces shown in the third row of Figure~\ref{fig:2D_noise_bipartite_18} provide a unified view of the joint dependence on noise rate and circuit depth. The resulting surfaces exhibit a transition-like geometry, consisting of relatively flat high-performance regions separated from rapidly degrading regions by a steep intermediate boundary. The geometry of these 3D surfaces suggests that the AR values may follow a logistic-like behaviour in three dimensions, with an intrinsic threshold separating high-performance and rapidly degrading regimes.

\begin{figure}[ht]
    \centering
    \includegraphics[width=0.95\linewidth]{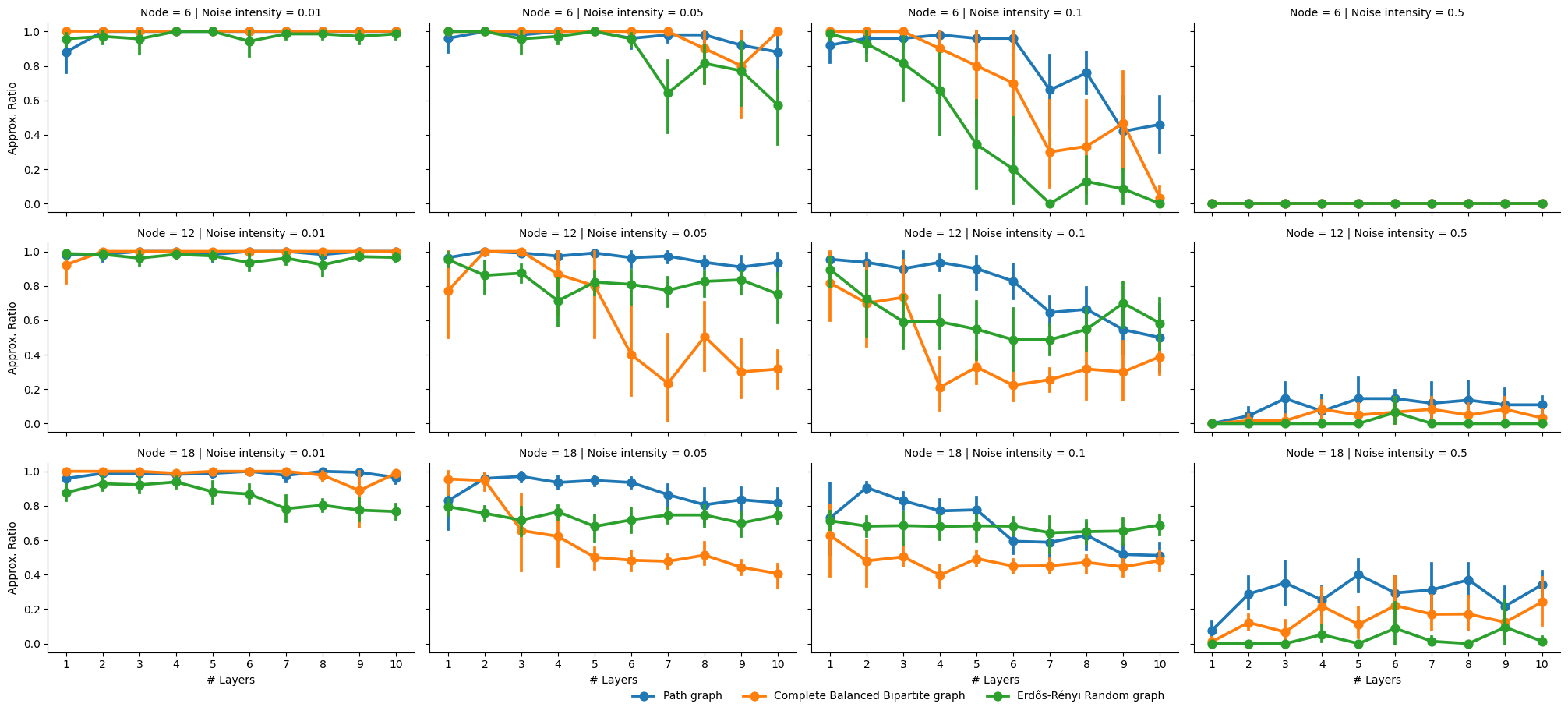}
    \caption{The evolution of QAOA approximation ratio (AR) with respect to circuit depth under the amplitude damping noise channel, compared across graph structure and graph order. Rows correspond to graph order ($n=6,12,18$ vertices) and columns correspond to noise intensity ($r=0.01,0.05,0.1,0.5$). Coloured curves show path graphs (blue), complete balanced bipartite graphs (orange) and Erd\H{o}s--R\'enyi random graphs (green); error bars denote the standard deviation across independent runs.}
    \label{fig:2D_instance_analysis_AmplitudeDamping}
\end{figure}

We further examine how the relationship between AR and the QAOA depth evolves across different graph structures and graph orders. Figure~\ref{fig:2D_instance_analysis_AmplitudeDamping} summarises representative results under the amplitude damping channel. 

Three characteristic regimes can be identified. In the low-noise regime, the AR remains close to optimal across all circuit depths, indicating that noise has negligible impact even on deeper circuits. In contrast, in the high-noise regime, the AR is uniformly low and largely insensitive to the number of layers, reflecting the dominance of noise over circuit expressivity. Between these two extremes lies a moderate-noise regime, where the behaviour becomes non-trivial: the AR exhibits a peak at a finite number of layers, suggesting the existence of an optimal choice of circuit depth.

Furthermore, the location of these regimes depends strongly on graph order. For example, at a noise rate of 0.05, graphs with six vertices considerably still behave similarly to the low-noise regime, whereas under the same noise intensity, larger graphs already exhibit transition-like behaviour with clearly identifiable optimal depths. This indicates that larger problem instances are increasingly sensitive to noise accumulation.

Graph structure further influences this behaviour. Sparse graphs, such as path graphs, generally maintain higher AR values across larger circuit depths. In contrast, denser graph structures exhibit earlier degradation due to the increased number of entangling two-qubit operations required by the corresponding QAOA circuits. Since two-qubit gates typically contribute the dominant source of hardware noise, denser graphs effectively induce greater exposure to accumulated errors.

Overall, these observations demonstrate that QAOA robustness depends not only on noise magnitude and circuit depth, but also on the instances' sizes and structural properties.

\subsection*{Optimal threshold modeling}

The transition-like behaviour observed in the 3D AR surfaces (the third row of Figure~\ref{fig:2D_noise_bipartite_18}) motivates the use of a logistic surface model to characterise the relationship between approximation ratio, circuit depth, and noise rate. In particular, we model the joint dependence of AR on these variables using a three-dimensional generalised logistic function capable of capturing both plateau regions and rapid degradation regions within a unified framework. The complete formulation of the fitting procedure and parameter estimation will be detailed in Section~\textit{Methods}. The resulting threshold is defined as the inflection boundary of the fitted logistic surface. Operationally, this corresponds to the region of maximal sensitivity, where small variations in noise rate produce the largest changes in AR. This provides a consistent quantitative definition of the boundary separating high-performance and noise-dominated regimes.

\begin{figure}[ht]
    \centering
    \includegraphics[width=0.95\linewidth]{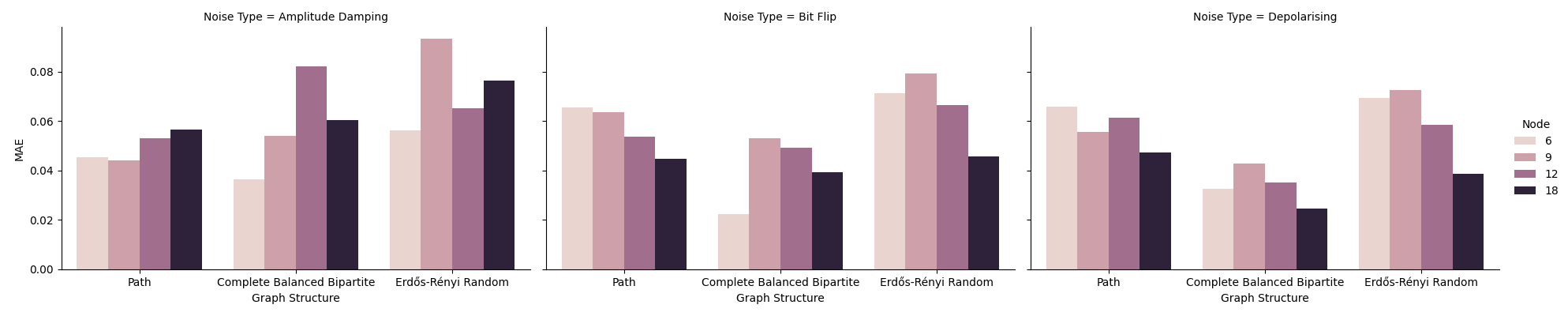}
    \caption{Accuracy of the fitted generalised logistic surface measured using the mean absolute error (MAE) across graph structures, graph orders and noise channels.}
    \label{fig:logistic_fit_mae}
\end{figure}
Figure~\ref{fig:logistic_fit_mae} summarises the accuracy measures of the logistic surface fitting using the mean absolute error (MAE) across different graph structures, noise models, and graph orders. The MAE metric quantifies the average absolute deviation between the fitted logistic surface and the observed AR values, meaning that lower MAE indicates a better fit. Overall, the logistic surface fitting demonstrates considerably low MAE (consistently less than 0.1) across most graph structures and noise models, indicating that the transition-like geometry of the AR surfaces is well captured by the proposed formulation.

\begin{figure}[ht]
    \centering
    \includegraphics[width=0.85\linewidth]{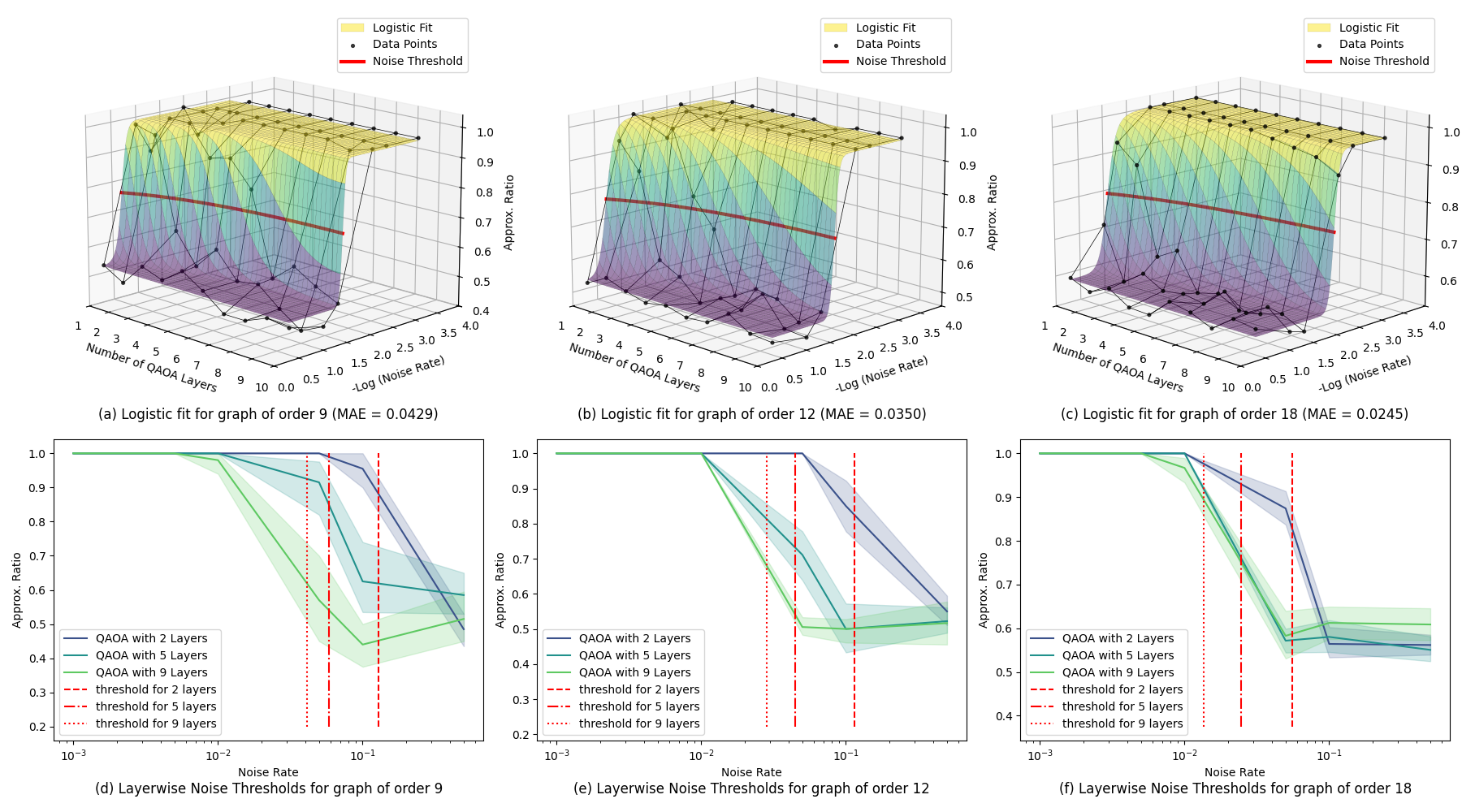}
    \caption{AR surface fitting using generalised logistic function and extracted noise thresholds for the path graph family under the depolarising channel. \textbf{(a–c)} Fitted generalised logistic surfaces (yellow-to-purple) overlaid on the observed approximation ratio data points (black), as a function of QAOA depth and $−\log(\text{noise rate})$, with the extracted noise threshold shown in red, for various graph orders. \textbf{(d–f)} Approximation ratio as a function of noise rate (log scale) for selected circuit depths ($p=2,5,9$), with vertical lines marking the corresponding layerwise noise thresholds extracted from the fitted surfaces.}
    \label{fig:logistic_fit_thresholds_snapshot}
\end{figure}

Complete results on the logistic modeling for all configurations are provided in Supplementary file Section B. Among the tested configurations, we select the path graph family under the depolarising channel as a representative snapshot for visualising the extracted threshold surfaces. Focusing on this specific case, Figure~\ref{fig:logistic_fit_thresholds_snapshot} presents the logistic modeling results applied to this case. The red curves indicate the noise thresholds extracted across different circuit depths. These threshold lines can be interpreted as phase boundaries separating regimes where QAOA maintains near-optimal performance from those where performance rapidly deteriorates. 

\begin{figure}[ht]
    \centering
    \includegraphics[width=0.8\linewidth]{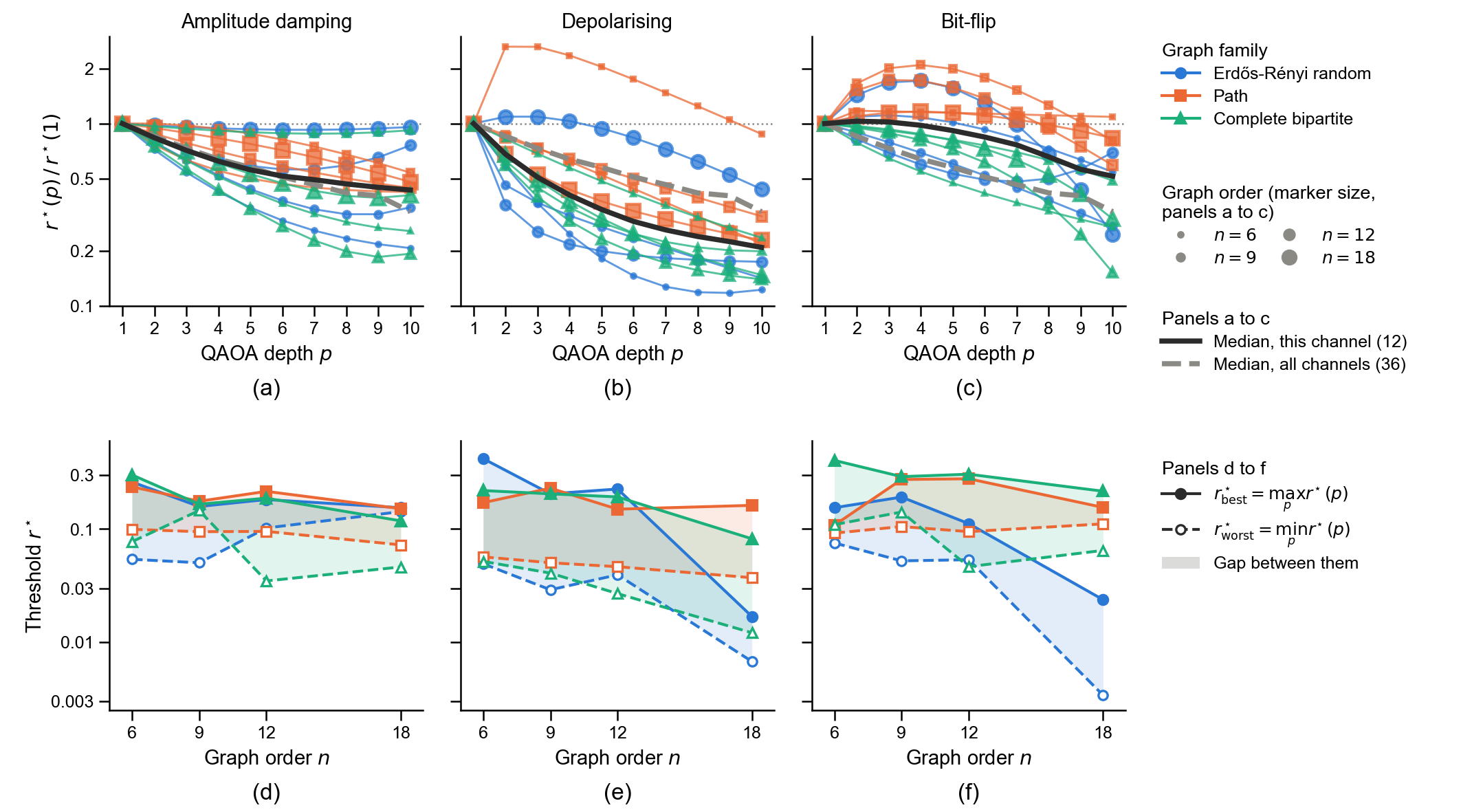}
    \caption{Dependence of the noise threshold on depth, noise channel and graph order. The columns correspond to three noise channels considered in this study. \textbf{(a)-(c)} Threshold relative to its single-layer value, $r^\star(p)/r^\star(1)$, as a function of QAOA depth $p$ for the 12 configurations of each channel. Colour and marker shape denote the graph family, and marker size denotes the graph order $n$. The thick solid line is the median over the 12 configurations of that channel, and the thick dashed line is the median over all 36 configurations. \textbf{(d)-(f)} Best-case threshold $r^\star_{\mathrm{best}}=\max\limits_{1\le p\le10}r^\star(p)$ (solid lines, filled markers) and worst-case threshold $r^\star_{\mathrm{worst}}=\min\limits_{1\le p\le10}r^\star(p)$ (dashed lines, open markers) as a function of graph order. The shaded band between them is the intermediate noise window. All vertical axes are logarithmic.}
    \label{fig:threshold_summary}
\end{figure}

In this example the threshold shifts towards lower noise rates as the QAOA depth increases. This means that deeper circuits require stricter noise conditions to maintain efficacy. The AR curves for different depths overlap below a common noise level and separate only above it, indicating that depth has little hindrance on the optimisation performance until this level is exceeded. Across graph orders, the threshold at $p=5$ decreases from 0.120 at $|V|=9$ to 0.085 at $|V|=12$ and 0.061 at $|V|=18$. Whether these trends hold more generally is examined in Figure~\ref{fig:threshold_summary}.

We now turn to how the threshold values varies with respect to QAOA depth, noise channel, and the graph instances, as illustrated in Figure~\ref{fig:threshold_summary}. Panels (a) to (c) trace the threshold relative to its single-layer value as a function of depth for each channel, and the panels (d) and (e) summarise its best and worst values over all depth choices as a function of graph order. 

Among the 36 combinations of channel, graph family, and graph order in panel (a)-(c), 35 cases have thresholds at $p=10$ that lie below those at $p=1$. The exception is the six-vertex path graph under bit-flip noise, for which the ratio is 1.09. The median ratio $r^\star(10)/r^\star(1)$ is 0.33 overall, but it depends on the channel (0.43 for amplitude damping, 0.21 for depolarising and 0.51 for bit-flip noise) and on the graph family (0.25 for complete bipartite, 0.31 for \ER{} random and 0.51 for path graphs). Hence, the decreasing trend of the threshold with depth is observed in the median and convey a meaningful indication: deeper QAOA circuits demand for stricter noise requirement to avoid noise-dominated AR degradation, generalising the phenomenon appearing in the snapshot example provided in Figure \ref{fig:logistic_fit_thresholds_snapshot}. 

The threshold declining fashion also depends on the channel. For amplitude damping and depolarising noise, the decline is steeper at shallower depths; in contrast to those of bit-flip channel whose steepest drops occur on the greater $p$. Moreover, individual trajectories are not always monotonic: 16 of the 36 rise over at least one step in depth. The most significant rise occurs under the six-vertex path graph under depolarising noise, whose threshold increases about 2.6-fold from $p=1$ to $p=2$. The stricter noise threshold appearing on smaller depths could indicate that the inexpressive shallower circuits might require reduced error rates in order to achieve its optimal performance.

Since the threshold trends differ between configurations, we summarise each threshold trajectory by its best-case threshold $r^\star_{\mathrm{best}}=\max\limits_{1\le p\le10}r^\star(p)$,  and its worst-case threshold $r^\star_{\mathrm{worst}}=\min\limits_{1\le p\le10}r^\star(p)$. These two values break the noise axis into three areas. Below $r^\star_{\mathrm{worst}}$, every depth up to $p=10$ remains in the high-performance region, so the conventional strategy of adding layers to improve the approximation ratio applies. Above $r^\star_{\mathrm{best}}$, on the other hand, no depth choice would be able to even generate an effective optimisation. In the intermediate window, success depends on choosing a sufficiently shallow depth in order to catch the most optimal optimisation result. The best case is attained at $p=1$ in 28 of the 36 configurations, and six of the eight exceptions occur under bit-flip noise, where it is attained between $p=2$ and $p=4$ whereas the worst case is attained at $p=10$ in 27 configurations. The ratio $r^\star_{\mathrm{best}}/r^\star_{\mathrm{worst}}$ has a median of 3.2 and ranges from 1.08 to 8.5, which signifies that the effective noise requirement could vary arbitrarily among channels and problem instances. When devising strategy for reliable QAOA, one could refer to this ratio value: finding for optimal number of QAOA layers might be more relevant to configurations where the ratio is bigger, especially when the employed device noise level falls within the intermediate window.

Both summary thresholds generally decrease with graph order, but not uniformly (Figure~\ref{fig:threshold_summary}d,e). The best-case threshold is lower at $|V|=18$ than at $|V|=6$ in 8 of the 9 channel-family groups. Meanwhile, the worst-case threshold also performs the same in 7 of the 9 configurations. The strongest dependence on order appears for \ER{} random graphs under bit-flip noise, whose worst-case threshold falls by a factor of about 22 between $|V|=6$ and $|V|=18$. Path graphs show the weakest dependence, and under bit-flip noise their worst-case threshold does not decrease at all. Complete bipartite graphs, on the other hand, show a consistent decrease under all three channels. This contrast is consistent with the noise exposure of each qubit: every vertex of a path graph takes part in at most $\mathbf{O}(1)$ two-qubit gates per layer regardless of $n$ whereas this number grows as $\mathbf{O}(n)$ for both complete bipartite and \ER{} random graphs. Consequently, an additional layer corrupts every qubit more as the graph grows. The noise threshold of QAOA thus appears to be governed less by the size of the problem than by its connectivity. Hence, an effective noise threshold cannot be stated independently of the noise channel and the graph structure.

\subsection*{Threshold-based assessment on a superconducting processor}

The notion of effective noise threshold is intended to serve as a pre-execution checking criterion: a depth-$p$ QAOA circuit is expected to remain in the high-performance regime on a device whose noise rate $r_{\mathrm{dev}}$ satisfies $r_{\mathrm{dev}}<r^\star(p)$. By construction of the fitted surface, the high-performance regime corresponds to an AR above the midpoint level of the surface, that is $y_0+L/2$ (see Methods), which is the AR at which the threshold boundary lies. To assess this, we applied this criterion to the \texttt{ibm\_marrakesh} superconducting processor, executing QAOA without error mitigation on the 18-vertex \ER{} instance for $1\le p\le10$ (see Methods). This experiment illustrates how the threshold would be used in practice. Due to computational constraints, examining a single problem instance and performing one execution per depth is the most feasible approach for the current situation. Hence, this discloses the experimental limitation, reserving full hardware validation as a primary objective for future research.

To estimate the value of $r_{\mathrm{dev}}$, we consider the publicly available device calibration data from the \texttt{FakeMarrakesh} backend as summarised in Table~\ref{tab:marrakesh_noise_stats}. We take the reported two-qubit gate error as the device noise rate $r_{\mathrm{dev}}$, whose median over the 163 usable couplers is 0.0032, with an interquartile range of 0.0024 to 0.0043. Single-qubit gate errors are about an order of magnitude smaller (median 0.00023), so the two-qubit gate error sets the noise level of the QAOA circuit. Because the simulated noise rate acts on each qubit separately, the two-qubit gate error overstates the corresponding per-qubit rate of the bit-flip and depolarising channels, which makes the comparison below conservative for these two channels.

\begin{table}[ht]
\centering
\caption{Calibration statistics of the \texttt{ibm\_marrakesh} gate errors, retrieved from \texttt{FakeMarrakesh} backend. }
\label{tab:marrakesh_noise_stats}
\begin{tabular}{lccccccc}
\hline
\textbf{Error type} & \textbf{Mean} & \textbf{Std} & \textbf{Min} & \textbf{25\%} & \textbf{50\%} & \textbf{75\%} & \textbf{Max} \\
\hline
Single-qubit gate error & 0.000858 & 0.003861 & 0.000069 & 0.000164 & 0.000230 & 0.000353 & 0.041946 \\
Two-qubit gate error & 0.007229 & 0.019670 & 0.001036 & 0.002437 & 0.003234 & 0.004253 & 0.163001 \\
\hline
\end{tabular}
\end{table}

\begin{figure}[ht]
\centering
\includegraphics[width=0.8\linewidth]{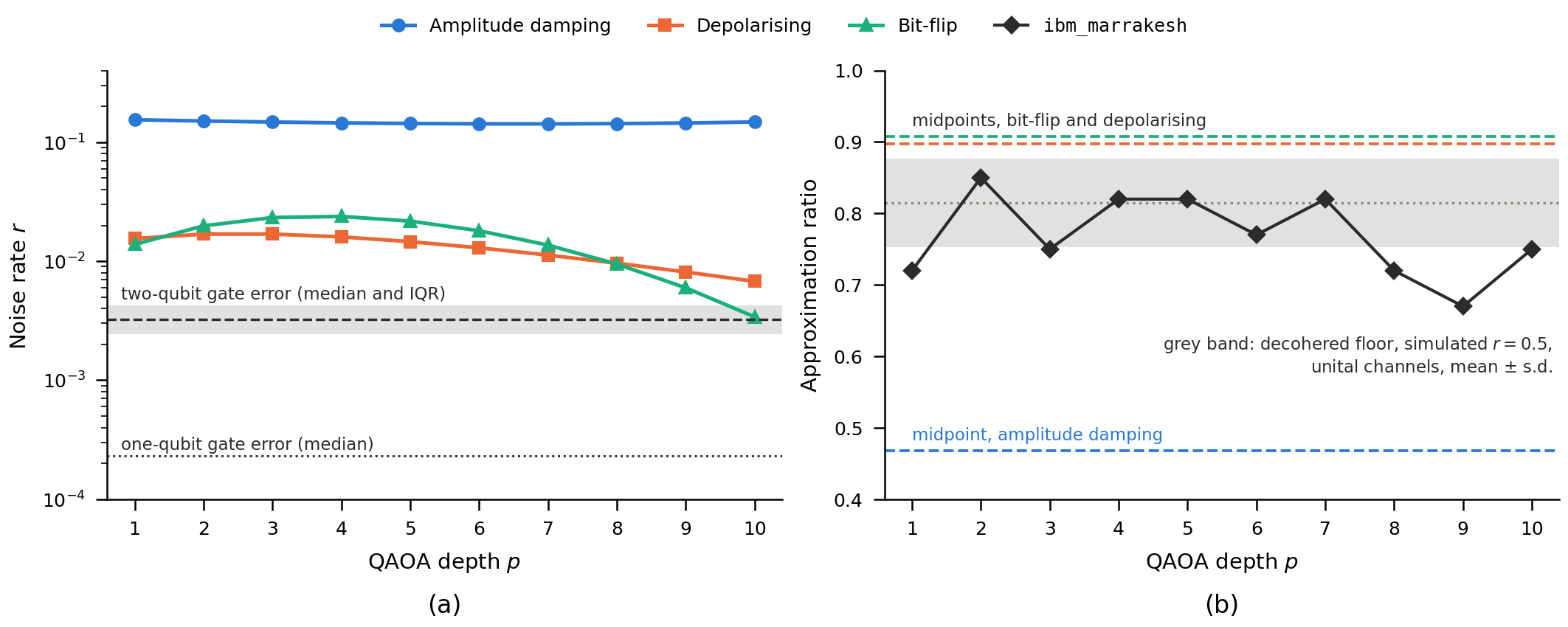}
\caption{Threshold-based assessment of QAOA on the \texttt{ibm\_marrakesh} processor for the 18-vertex \ER{} instance. \textbf{(a)} Effective noise thresholds $r^\star(p)$ under the three noise channels, compared with the device noise rate. The grey band and dashed line show the interquartile range and median of the reported two-qubit gate error, and the dotted line shows the median single-qubit gate error. \textbf{(b)} Approximation ratio obtained on hardware at each depth. Dashed lines mark the midpoint AR of the fitted logistic surface for each channel, which is the AR at the threshold. The grey band shows the mean $\pm$ standard deviation of the simulated AR at $r=0.5$ under the unital channels, which approximates the AR returned by a fully decohered register.}
\label{fig:quantum_threshold}
\end{figure}

Figure~\ref{fig:quantum_threshold}a places the device rate against the thresholds of the tested instance. We observe that the 2-qubit median error rate taken as $r_{\mathrm{dev}}$ ($0.003234$) lies below all the worst-case threshold $r^\star_{\mathrm{worst}}$ in all 36 configurations. Nevertheless, there is a single case, that is the 18-vertex \ER{} instance under bit-flip noise, where the worst threshold $r^\star_{\mathrm{worst}}=0.0034$ lies nearly  $r_{\mathrm{dev}}$. This also motivates the problem instance choice for the hardware runs. For this instance, the amplitude damping thresholds substantially exceed $r_{\mathrm{dev}}$  at every depth, whereas the bit-flip and depolarising thresholds decline with depth and approach $r_{\mathrm{dev}}$ only at the largest depths. The smallest margin occurs under bit-flip noise at $p=10$. 

The hardware results are shown in Figure~\ref{fig:quantum_threshold}b. The AR obtained on the device ranges from 0.67 to 0.85 with a mean of 0.77. Accompanying the device AR plot, the midpoint levels of the bit-flip (0.907), depolarising (0.897) and amplitude damping (0.468) surface are provided as the dashed lines coloured accordingly to indicate their corresponding channels. As a reference for a register that retains no useful coherence, the simulated AR at $r=0.5$ under the unital channels (depolarising and bit-flip combined) is $0.815\pm0.062$. This particular observation will be useful for the upcoming discussion. AR values under such channels are far from zero because the reported solution is the most frequent of 16,384 samples drawn from $2^{18}$ possible bitstrings, with ties broken towards the larger cut, which amounts to selecting the best of many near-random samples. Five of the ten hardware values lie within one standard deviation of this floor.  

Although the hardware results might not fit exactly as a quantitative test of the thresholds, the threshold modeling itself is still beneficial for interpreting them. It is important to note that the executed hardware circuits were degraded by elements that the model does not include, each of which lowers the AR relative to that of the logical circuits from which the modelled thresholds are derived. Hence, the hardware AR should not be expected to match the simulated AR levels in absolute terms as this would not serve as the goal of this hardware assessment either. For example, readout errors act on all 18 qubits at the end of every circuit, and their effect is approximately independent of depth. Degrading elements that scale up with depth also exist. Routing the dense \ER{} instance onto the heavy-hex connectivity of the device also introduces additional two-qubit SWAP gates in every cost layer. In addition, the optimisation budget on hardware was 125 COBYLA iterations against 250 in simulation (see Methods). Both of these effects grow with depth. Two features of the hardware AR nevertheless remain informative from the corresponding noise model: the AR relative position, which indicates the type of noise channel dominating the device; and its variation with depth, which can be compared with the depth dependence of the thresholds.

In light of the threshold modeling results, the assessment of the hardware AR relative position indicates that the device noise contains a substantial relaxation-type contribution. To see this, we start by observing that at every depth, the hardware AR lies below the midpoint levels of both unital surfaces. At first, this might pose a deviation from what the threshold model initially suggests, but a closer look would find this phenomenon informative. The decohered floor in Fig.~\ref{fig:quantum_threshold}b clarifies this: at $p=1$, $8$ and $9$, the hardware AR falls below the band, that is, below the AR typically returned by a register that the unital channels have fully decohered. As these channels drive the register towards the maximally mixed state, it is important to observe that even the operationally strongest unital noise does not degrade the AR as bad as the physical noisy environment of the hardware does. This implies that the unital channels alone cannot account for the device operational behaviour. Instead, the amplitude damping can moderately fill this gap as it drives the state towards $|0\rangle^{\otimes n}$, whose cut value is zero. In fact, the hardware AR remains above the midpoint level of this channel at every depth, in line with its threshold exceeding $r_{\mathrm{dev}}$ throughout. This picture is consistent with the prominence of $T_1$ decay in superconducting qubits and with the asymmetry of their readout errors, which favour the transition from $|1\rangle$ to $|0\rangle$\cite{chen2023transmon}.

Although the hardware noise is less contributed by unital channels, the depth dependence of the hardware AR still follows the unital channels' thresholds. Under the bit-flip and depolarising channels, the thresholds decline beyond $p\approx4$ and approach $r_{\mathrm{dev}}$ at the highest depths. Evidently, the hardware AR is lowest in the similar range, with a mean of 0.79 for $p\le7$ and 0.71 for $p\ge8$. The amplitude damping threshold, by contrast, is nearly independent of depth and remains far above $r_{\mathrm{dev}}$, so this channel identifies no specific depth at which the device approaches its threshold. The two observations therefore assign complementary roles to the channel models: amplitude damping accounts for the overall relative level of the hardware AR, whereas the unital channels capture its variation with depth through the proximity of their thresholds to $r_{\mathrm{dev}}$.

Finally, this simple hardware assessment identifies an opportunity to investigate device-specific thresholds. Since the hardware exhibits features of both unital and non-unital noise, the threshold extraction would benefit from simulations with a composite noise model whose components and rates are informed from the calibration of the target device, applied to the transpiled circuit and including readout error. In this setting, $r^\star$ would become a boundary in a multi-parameter noise space rather than a single rate. Repeated executions at each depth, several problem instances examination, and choosing a specific instance native to the device qubits connectivity, would allow fairer structural correspondence to be tested.

\section*{Discussion}

We have successfully demonstrated characterisation of noisy QAOA not by the depth that performs best at a given noise rate, but by the noise rate that each depth can tolerate. Across three noise channels, three graph families and graphs of up to 18 vertices, the approximation ratio remains high up to a noise level corresponding to the effective noise threshold and then degrades abruptly, by utilising a geometry that a generalised logistic surface captures with a mean absolute error below 0.1 in most configurations. The modeled noise threshold $r^\star(p)$, defined as the inflection boundary of this surface, declines with depth in 35 of the 36 configurations, with a median $r^\star(10)/r^\star(1)$ of 0.33. 

The best-case and worst-case thresholds, $r^\star_{\mathrm{best}}$ and $r^\star_{\mathrm{worst}}$, divide the noise axis into three regions that call for different strategies. Below $r^\star_{\mathrm{worst}}$, every depth up to $p=10$ contribute to a reliable approximation, meaning that the predicted AR remains above the midpoint of the fitted surface. In this situation, QAOA depth can be chosen on grounds of expressivity, as in the noiseless setting. Above $r^\star_{\mathrm{best}}$, no depth within the range induces effective optimisation. Here, resources are better directed towards error mitigation\cite{chi2026variational}, sparser problem encodings\cite{wang2025depth} or noise-informed hamiltonian remapping\cite{maciejewski2025improving} than towards depth selection. Between the two lies a window in which the outcome depends on choosing a sufficiently shallow circuit. The width between these two, measured by $r^\star_{\mathrm{best}}/r^\star_{\mathrm{worst}}$, has a median of 3.2 and ranges from 1.08 to 8.5 across configurations. These bands constitute an area where QAOA depth selection strategies\cite{zhou2020quantum,pan2022efficient,saini2025dynamic,skolik2021layerwise} play the most crucial role to obtain reliable QAOA results, as the interplay between circuit expressivity and noise domination displays non trivial relationship here. These strategies are therefore complementary to the threshold perspective and most valuable inside the window, particularly where $r^\star_{\mathrm{best}}/r^\star_{\mathrm{worst}}$ is wide. Notably, the noise rate in this context should be understood as the modelled rate governed faithfully by the noise channel. The hardware results, however, show that the factual noise rate can exceed the modelled rate.

The threshold perspective also suggests why a simpler empirical description of the same dependence had limited success. Liu et al.~\cite{liu2025noisy} fitted linear models of the approximation ratio against graph size, edge count and noise rate, over noise rates between 0.1 and 0.9. Their models found negligible explanatory power for complete graphs ($R^2=0.0005$) and moderate power for random graphs ($R^2=0.67$). Their noise is applied once per layer rather than after each gate, so their rates are not directly comparable with ours. Nevertheless, a linear predictor fitted largely over the saturated part of a sigmoidal response would be expected to show little dependence on its inputs, and our results indicate that the approximation ratio responds to noise in this sigmoidal manner. The generalised logistic surface used here can be read as a linear predictor passed through a bounded sigmoid link and extended with the circuit depth factor and quadratic interaction terms, so that the transition itself is represented in the model rather than averaged across. Additionally, since the surface yields a threshold for each depth, it also supports a check before execution in either direction: 1) given a device, which depths and problem instances are expected to remain effective, and 2) given a problem instance, what noise level a device must reach for a chosen depth.

The dependence on the noise model is the main lesson of the hardware assessment. The median two-qubit gate error of \texttt{ibm\_marrakesh} lies below the worst-case threshold of every simulated configuration, so a check based on this error alone would predict an AR above the midpoint level of each channel at every depth. The hardware AR, with a mean of 0.77, instead lies below the midpoint levels of both unital channels and, at several depths, below the AR returned by a register that these channels have fully decohered. The modeled gate error therefore understates the factual noise rate of the hardware circuit, which also carries routing overhead and readout error. Moreover it is also subject to the relaxation that the unital channels cannot represent. These individual channel models nevertheless remain informative: amplitude damping accounts for the overall relative level of the hardware AR, whereas its depth dependence follows the unital thresholds. With a single problem instance, a single execution per depth and a single calibration snapshot, the experiment illustrates how the threshold would be applied in practice rather than validating it. Closing the gap requires thresholds extracted from a device-specific noise modeling strategy. Simulations with a composite noise model informed by its calibration, combining various noise sources, and applied to the hardware-aware transpiled circuit; would turn $r^\star$ into a boundary in a multi-parameter noise space. Subsequent investigations employing factorial linear analysis\cite{natanael2026exhaustive} to evaluate element-specific noise effects could provide a deeper insight  into the dynamics of error propagation within variational quantum circuits.

More broadly, we observe that the effective noise threshold can serve as a generic metric for assessing the noise resilience of variational quantum algorithms, highlighting several promising directions for future noise-related research. Many studies of noisy QAOA report performance at a few fixed noise rates~\cite{pellow2024effect,lotshaw2021empirical,xue2021effects}, which makes results difficult to compare across problem classes and the device resources. Sweeping the noise rate and extracting $r^\star_{\mathrm{best}}$ and $r^\star_{\mathrm{worst}}$ instead yields comparable metrics across other factors. Subsequently, a threshold computed for a certain targeted hardware workload translates directly into an error-rate requirement, that provides a workload-specific metric of merit alongside generic measures such as gate fidelity. Finally, whether the transition-like degradation observed here is transferrable into a more general feature of noisy variational optimisation algorithms across arbitrary problems remains an open question.

\section*{Methods}

\subsection*{Generalised Logistic Curve Fitting for Noise Threshold Estimation}

To quantitatively characterise the transition between high-performing and degraded QAOA regimes under increasing noise, we model the approximation ratio (AR) 3D surface using a generalised logistic function. The logistic formulation is motivated by the empirical observation that QAOA performance degradation under noise often exhibits a smooth transition behaviour resembling an s-shaped curve. The objective of this modeling step is to estimate the \emph{effective noise threshold}, approximated to be around the inflection point at which the fitted logistic surface begins to undergo rapid transition. In this context, the inflection point of the logistic surface corresponds to the critical boundary separating relatively robust and strongly degraded operating regimes. Specifically, the generalised logistic model is defined as follows:
\[
f_{y_0,L,b,\boldsymbol{\theta}}(\mathbf{x})
=
y_0+
\frac{L}
{1+\exp\!\left(-b-\mathbf{x}\cdot\boldsymbol{\theta}\right)},
\]
where $y_0,L,b\in\mathbb{R}$ are scalar parameters and
$\boldsymbol{\theta}\in\mathbb{R}^d$ is the parameter vector associated with the feature vector $\mathbf{x}$. The inflection point occurs when the exponential argument equals 0, that is $b+\mathbf{x}\cdot\boldsymbol{\theta}=0$.
This condition is therefore employed to estimate the corresponding noise threshold.

\subsubsection*{Feature Construction}

To capture nonlinear interactions between circuit depth and noise level, we augment the feature representation with auxiliary quadratic terms. Let $p$ and $r$ be the QAOA depth and the noise rate respectively. The noise rate is first transformed to $s=-\log_{10}r$, which maps the logarithmically spaced noise rates onto a uniform scale and improves numerical conditioning. To prepare for a good fitting quality, both variables are standardised: $p'=(p-\mu_p)/\sigma_p$ and $s'=(s-\mu_s)/\sigma_s$, where the means and standard deviations are computed over the experimental grid and are common to all fits. The model predicts the approximation ratio from the following augmented features: $\textbf{x}=( p',s', p'^2,s'^2, p's')$. Hence, the logistic decision boundary becomes
\begin{equation}b+\theta_1 p'+\theta_2s'+\theta_3 p'^2+\theta_4s'^2+\theta_5 p's'=0
\label{quadratic_eq_ori}\end{equation}
which in general would appear as a two-dimensional quadratic curve.

\subsubsection*{Threshold Extraction for Fixed Circuit Depth}
To estimate the noise threshold for a fixed QAOA depth $p$, we solve the decision boundary for $s'$. Rearranging equation \eqref{quadratic_eq_ori} gives the following the quadratic equation in ${s'}$
\begin{equation}\theta_4s'^2+(\theta_2+\theta_5 p')s'+(b+\theta_1 p'+\theta_3 p'^2)=0,\end{equation}
whose roots are
\begin{equation}s'_p=\frac{-(\theta_2+\theta_5 p')\pm\sqrt{(\theta_2+\theta_5 p')^2-4\theta_4(b+\theta_1 p'+\theta_3p'^2)}}{2\theta_4}.\end{equation}
The threshold noise rate is recovered by the inverse transformation $r_p=10^{-(\sigma_ss'_p+\mu_s)}$. Only real roots are retained, and among them we pick only the root lying within the sampled range of $s$. When no such root exists, no threshold is reported.

\subsection*{Noisy Simulation Implementation Setup}
All noisy simulations were implemented using the NVIDIA CUDA-Q framework\cite{kim2023cuda}. Noise was incorporated into both the optimisation and evaluation stages of the QAOA workflow via CUDA-Q's noise-model interface. During classical optimisation, the expectation value of the problem Hamiltonian was estimated using \texttt{cudaq.observe()} with the noise model applied, using 1024 stochastic trajectories per evaluation to approximate the noisy expectation value. Parameter optimisation was performed using the COBYLA optimiser with a maximum of 250 iterations.

Following optimisation, the variational parameters corresponding to the iteration that achieved the lowest expectation energy were selected for final evaluation. Note that the ultimately selected circuit might not be the one produced by the last optimisation iteration. The final circuit was then sampled under the noise model using \texttt{cudaq.sample()} with 16384 measurement shots. The resulting bitstring distribution was ranked by measurement count in descending order, with ties broken by preferring the bitstring corresponding to the larger MaxCut value.

\subsection*{Real Quantum Hardware Implementation}

In this section, we describe the implementation methodology used for executing QAOA on real quantum hardware. All hardware experiments were conducted using the \texttt{ibm\_marrakesh} backend available through the IBM Quantum platform. The QAOA parameter optimisation process on quantum hardware employed the COBYLA classical optimiser. To maintain feasible execution time and hardware resource usage, the maximum number of COBYLA iterations was restricted to \texttt{maxiter}=125. During each optimisation iteration, expectation values of the problem Hamiltonian were estimated using the IBM Quantum \texttt{EstimatorV2} primitive. Each estimator evaluation used 4096 measurement shots to reduce statistical fluctuations while keeping the total hardware execution cost manageable. After completion of the optimisation process, the final QAOA circuit was evaluated using the IBM Quantum \texttt{SamplerV2} primitive. In this stage, 16,384 shots was employed to obtain a more reliable approximation of the final output probability distribution.

To preserve the authenticity of the observed hardware behaviour, no error mitigation or noise suppression techniques were applied throughout the experiment. This choice allows the reported results to more directly reflect the intrinsic noise characteristics of current NISQ hardware. Prior to hardware execution, all circuits were transpiled using the Qiskit transpiler with optimisation level 3, which corresponds to the highest built-in optimisation setting available in the standard transpilation workflow. This optimisation stage attempts to reduce circuit depth, minimise two-qubit gate count, and improve qubit layout selection according to the calibration properties of the target backend.

\bibliography{nature}

\begin{thebibliography}{10}
\urlstyle{rm}
\expandafter\ifx\csname url\endcsname\relax
  \def\url#1{\texttt{#1}}\fi
\expandafter\ifx\csname urlprefix\endcsname\relax\def\urlprefix{URL }\fi
\expandafter\ifx\csname doiprefix\endcsname\relax\def\doiprefix{DOI: }\fi
\providecommand{\bibinfo}[2]{#2}
\providecommand{\eprint}[2][]{\url{#2}}

\bibitem{abbas2024challenges}
\bibinfo{author}{Abbas, A.} \emph{et~al.}
\newblock \bibinfo{journal}{\bibinfo{title}{Challenges and opportunities in
  quantum optimization}}.
\newblock {\emph{\JournalTitle{Nature Reviews Physics}}}
  \textbf{\bibinfo{volume}{6}}, \bibinfo{pages}{718--735}
  (\bibinfo{year}{2024}).

\bibitem{pirnay2024principle}
\bibinfo{author}{Pirnay, N.}, \bibinfo{author}{Ulitzsch, V.},
  \bibinfo{author}{Wilde, F.}, \bibinfo{author}{Eisert, J.} \&
  \bibinfo{author}{Seifert, J.-P.}
\newblock \bibinfo{journal}{\bibinfo{title}{An in-principle super-polynomial
  quantum advantage for approximating combinatorial optimization problems via
  computational learning theory}}.
\newblock {\emph{\JournalTitle{Science advances}}}
  \textbf{\bibinfo{volume}{10}}, \bibinfo{pages}{eadj5170}
  (\bibinfo{year}{2024}).

\bibitem{moncayo2025quantum}
\bibinfo{author}{Moncayo-Mart{\'\i}nez, L.~A.} \& \bibinfo{author}{He, N.}
\newblock \bibinfo{journal}{\bibinfo{title}{Quantum optimisation for supply
  chain: Qubo formulations and qaoa solutions for facility location and load
  balancing}}.
\newblock {\emph{\JournalTitle{Results in Engineering}}}
  \bibinfo{pages}{108373} (\bibinfo{year}{2025}).

\bibitem{yuan2026quantifying}
\bibinfo{author}{Yuan, H.}, \bibinfo{author}{Long, C.~K.},
  \bibinfo{author}{Lepage, H.~V.} \& \bibinfo{author}{Barnes, C.~H.}
\newblock \bibinfo{journal}{\bibinfo{title}{Quantifying the advantages of
  applying quantum approximate algorithms to portfolio optimisation}}.
\newblock {\emph{\JournalTitle{Quantum Science and Technology}}}
  \textbf{\bibinfo{volume}{11}}, \bibinfo{pages}{025034}
  (\bibinfo{year}{2026}).

\bibitem{yeo2025quantum}
\bibinfo{author}{Yeo, S.~S.} \& \bibinfo{author}{Han, J.~M.}
\newblock \bibinfo{title}{Quantum approximate optimisation algorithm (qaoa) for
  surplus energy distribution in urban microgrids}.
\newblock In \emph{\bibinfo{booktitle}{19th IBPSA Conference on Building
  Simulation, BS 2025}} (\bibinfo{organization}{International Building
  Performance Simulation Association}, \bibinfo{year}{2025}).

\bibitem{albash2018adiabatic}
\bibinfo{author}{Albash, T.} \& \bibinfo{author}{Lidar, D.~A.}
\newblock \bibinfo{journal}{\bibinfo{title}{Adiabatic quantum computation}}.
\newblock {\emph{\JournalTitle{Reviews of Modern Physics}}}
  \textbf{\bibinfo{volume}{90}}, \bibinfo{pages}{015002}
  (\bibinfo{year}{2018}).

\bibitem{cerezo2021variational}
\bibinfo{author}{Cerezo, M.} \emph{et~al.}
\newblock \bibinfo{journal}{\bibinfo{title}{Variational quantum algorithms}}.
\newblock {\emph{\JournalTitle{Nature Reviews Physics}}}
  \textbf{\bibinfo{volume}{3}}, \bibinfo{pages}{625--644}
  (\bibinfo{year}{2021}).

\bibitem{farhi2014quantum}
\bibinfo{author}{Farhi, E.}, \bibinfo{author}{Goldstone, J.} \&
  \bibinfo{author}{Gutmann, S.}
\newblock \bibinfo{journal}{\bibinfo{title}{A quantum approximate optimization
  algorithm}}.
\newblock {\emph{\JournalTitle{arXiv preprint arXiv:1411.4028}}}
  (\bibinfo{year}{2014}).

\bibitem{li2026auxiliary}
\bibinfo{author}{Li, G.} \emph{et~al.}
\newblock \bibinfo{journal}{\bibinfo{title}{Auxiliary-qubit-free quantum
  approximate optimization algorithm for the minimum dominating set problem}}.
\newblock {\emph{\JournalTitle{Chinese Physics B}}}
  \textbf{\bibinfo{volume}{35}}, \bibinfo{pages}{050304}
  (\bibinfo{year}{2026}).

\bibitem{pellow2024effect}
\bibinfo{author}{Pellow-Jarman, A.} \emph{et~al.}
\newblock \bibinfo{journal}{\bibinfo{title}{The effect of classical optimizers
  and ansatz depth on qaoa performance in noisy devices}}.
\newblock {\emph{\JournalTitle{Scientific reports}}}
  \textbf{\bibinfo{volume}{14}}, \bibinfo{pages}{16011} (\bibinfo{year}{2024}).

\bibitem{liu2025noisy}
\bibinfo{author}{Liu, D.}, \bibinfo{author}{Li, J.}, \bibinfo{author}{Chen, X.}
  \& \bibinfo{author}{Jiang, N.}
\newblock \bibinfo{journal}{\bibinfo{title}{Noisy quantum approximation
  optimization algorithm for solving maxcut problem}}.
\newblock {\emph{\JournalTitle{Physica Scripta}}}
  \textbf{\bibinfo{volume}{100}}, \bibinfo{pages}{096009}
  (\bibinfo{year}{2025}).

\bibitem{harrigan2021quantum}
\bibinfo{author}{Harrigan, M.~P.} \emph{et~al.}
\newblock \bibinfo{journal}{\bibinfo{title}{Quantum approximate optimization of
  non-planar graph problems on a planar superconducting processor}}.
\newblock {\emph{\JournalTitle{Nature Physics}}} \textbf{\bibinfo{volume}{17}},
  \bibinfo{pages}{332--336} (\bibinfo{year}{2021}).

\bibitem{weidenfeller2022scaling}
\bibinfo{author}{Weidenfeller, J.} \emph{et~al.}
\newblock \bibinfo{journal}{\bibinfo{title}{Scaling of the quantum approximate
  optimization algorithm on superconducting qubit based hardware}}.
\newblock {\emph{\JournalTitle{Quantum}}} \textbf{\bibinfo{volume}{6}},
  \bibinfo{pages}{870} (\bibinfo{year}{2022}).

\bibitem{suksmono2025quantum}
\bibinfo{author}{Suksmono, A.~B.}
\newblock \bibinfo{journal}{\bibinfo{title}{A quantum approximate optimization
  method for finding hadamard matrices}}.
\newblock {\emph{\JournalTitle{Scientific Reports}}}
  \textbf{\bibinfo{volume}{15}}, \bibinfo{pages}{33254} (\bibinfo{year}{2025}).

\bibitem{ebadi2022quantum}
\bibinfo{author}{Ebadi, S.} \emph{et~al.}
\newblock \bibinfo{journal}{\bibinfo{title}{Quantum optimization of maximum
  independent set using rydberg atom arrays}}.
\newblock {\emph{\JournalTitle{Science}}} \textbf{\bibinfo{volume}{376}},
  \bibinfo{pages}{1209--1215} (\bibinfo{year}{2022}).

\bibitem{thelen2024approximating}
\bibinfo{author}{Thelen, S.}, \bibinfo{author}{Safi, H.} \&
  \bibinfo{author}{Mauerer, W.}
\newblock \bibinfo{title}{Approximating under the influence of quantum noise
  and compute power}.
\newblock In \emph{\bibinfo{booktitle}{2024 IEEE International Conference on
  Quantum Computing and Engineering (QCE)}}, vol.~\bibinfo{volume}{2},
  \bibinfo{pages}{274--279} (\bibinfo{organization}{IEEE},
  \bibinfo{year}{2024}).

\bibitem{marshall2020characterizing}
\bibinfo{author}{Marshall, J.}, \bibinfo{author}{Wudarski, F.},
  \bibinfo{author}{Hadfield, S.} \& \bibinfo{author}{Hogg, T.}
\newblock \bibinfo{journal}{\bibinfo{title}{Characterizing local noise in qaoa
  circuits}}.
\newblock {\emph{\JournalTitle{Iop Scinotes}}} \textbf{\bibinfo{volume}{1}},
  \bibinfo{pages}{025208} (\bibinfo{year}{2020}).

\bibitem{franca2021limitations}
\bibinfo{author}{Stilck~Fran{\c{c}}a, D.} \&
  \bibinfo{author}{Garc{\'\i}a-Patr{\'o}n, R.}
\newblock \bibinfo{journal}{\bibinfo{title}{Limitations of optimization
  algorithms on noisy quantum devices}}.
\newblock {\emph{\JournalTitle{Nature Physics}}} \textbf{\bibinfo{volume}{17}},
  \bibinfo{pages}{1221--1227} (\bibinfo{year}{2021}).

\bibitem{wang2021noise}
\bibinfo{author}{Wang, S.} \emph{et~al.}
\newblock \bibinfo{journal}{\bibinfo{title}{Noise-induced barren plateaus in
  variational quantum algorithms}}.
\newblock {\emph{\JournalTitle{Nature communications}}}
  \textbf{\bibinfo{volume}{12}}, \bibinfo{pages}{6961} (\bibinfo{year}{2021}).

\bibitem{mele2026noise}
\bibinfo{author}{Mele, A.~A.} \emph{et~al.}
\newblock \bibinfo{journal}{\bibinfo{title}{Noise-induced shallow circuits and
  the absence of barren plateaus}}.
\newblock {\emph{\JournalTitle{Nature Physics}}} \bibinfo{pages}{1--6}
  (\bibinfo{year}{2026}).

\bibitem{zhou2020quantum}
\bibinfo{author}{Zhou, L.}, \bibinfo{author}{Wang, S.-T.},
  \bibinfo{author}{Choi, S.}, \bibinfo{author}{Pichler, H.} \&
  \bibinfo{author}{Lukin, M.~D.}
\newblock \bibinfo{journal}{\bibinfo{title}{Quantum approximate optimization
  algorithm: Performance, mechanism, and implementation on near-term devices}}.
\newblock {\emph{\JournalTitle{Physical Review X}}}
  \textbf{\bibinfo{volume}{10}}, \bibinfo{pages}{021067}
  (\bibinfo{year}{2020}).

\bibitem{zhu2022adaptive}
\bibinfo{author}{Zhu, L.} \emph{et~al.}
\newblock \bibinfo{journal}{\bibinfo{title}{An adaptive quantum approximate
  optimization algorithm for solving combinatorial problems on a quantum
  computer}}.
\newblock {\emph{\JournalTitle{Physical Review Research}}}
  \textbf{\bibinfo{volume}{4}}, \bibinfo{pages}{033029} (\bibinfo{year}{2022}).

\bibitem{pan2022efficient}
\bibinfo{author}{Pan, Y.}, \bibinfo{author}{Tong, Y.}, \bibinfo{author}{Xue,
  S.} \& \bibinfo{author}{Zhang, G.}
\newblock \bibinfo{journal}{\bibinfo{title}{Efficient depth selection for the
  implementation of noisy quantum approximate optimization algorithm}}.
\newblock {\emph{\JournalTitle{Journal of the Franklin Institute}}}
  \textbf{\bibinfo{volume}{359}}, \bibinfo{pages}{11273--11287}
  (\bibinfo{year}{2022}).

\bibitem{saini2025dynamic}
\bibinfo{author}{Saini, R.}, \bibinfo{author}{Mohamed, N.},
  \bibinfo{author}{Al-Kuwari, S.} \& \bibinfo{author}{Farouk, A.}
\newblock \bibinfo{journal}{\bibinfo{title}{Dynamic depth quantum approximate
  optimization algorithm for solving constrained shortest path problem}}.
\newblock {\emph{\JournalTitle{arXiv preprint arXiv:2511.08657}}}
  (\bibinfo{year}{2025}).

\bibitem{skolik2021layerwise}
\bibinfo{author}{Skolik, A.}, \bibinfo{author}{McClean, J.~R.},
  \bibinfo{author}{Mohseni, M.}, \bibinfo{author}{van~der Smagt, P.} \&
  \bibinfo{author}{Leib, M.}
\newblock \bibinfo{journal}{\bibinfo{title}{Layerwise learning for quantum
  neural networks}}.
\newblock {\emph{\JournalTitle{Quantum Machine Intelligence}}}
  \textbf{\bibinfo{volume}{3}}, \bibinfo{pages}{5} (\bibinfo{year}{2021}).

\bibitem{zhang2022differentiable}
\bibinfo{author}{Zhang, S.-X.}, \bibinfo{author}{Hsieh, C.-Y.},
  \bibinfo{author}{Zhang, S.} \& \bibinfo{author}{Yao, H.}
\newblock \bibinfo{journal}{\bibinfo{title}{Differentiable quantum architecture
  search}}.
\newblock {\emph{\JournalTitle{Quantum Science \& Technology}}}
  \textbf{\bibinfo{volume}{7}}, \bibinfo{pages}{045023} (\bibinfo{year}{2022}).

\bibitem{du2022quantum}
\bibinfo{author}{Du, Y.}, \bibinfo{author}{Huang, T.}, \bibinfo{author}{You,
  S.}, \bibinfo{author}{Hsieh, M.-H.} \& \bibinfo{author}{Tao, D.}
\newblock \bibinfo{journal}{\bibinfo{title}{Quantum circuit architecture search
  for variational quantum algorithms}}.
\newblock {\emph{\JournalTitle{npj Quantum Information}}}
  \textbf{\bibinfo{volume}{8}}, \bibinfo{pages}{62} (\bibinfo{year}{2022}).

\bibitem{lotshaw2021empirical}
\bibinfo{author}{Lotshaw, P.~C.}, \bibinfo{author}{Humble, T.~S.},
  \bibinfo{author}{Herrman, R.}, \bibinfo{author}{Ostrowski, J.} \&
  \bibinfo{author}{Siopsis, G.}
\newblock \bibinfo{journal}{\bibinfo{title}{Empirical performance bounds for
  quantum approximate optimization}}.
\newblock {\emph{\JournalTitle{Quantum Information Processing}}}
  \textbf{\bibinfo{volume}{20}}, \bibinfo{pages}{403} (\bibinfo{year}{2021}).

\bibitem{xue2021effects}
\bibinfo{author}{Xue, C.}, \bibinfo{author}{Chen, Z.-Y.}, \bibinfo{author}{Wu,
  Y.-C.} \& \bibinfo{author}{Guo, G.-P.}
\newblock \bibinfo{journal}{\bibinfo{title}{Effects of quantum noise on quantum
  approximate optimization algorithm}}.
\newblock {\emph{\JournalTitle{Chinese Physics Letters}}}
  \textbf{\bibinfo{volume}{38}}, \bibinfo{pages}{030302}
  (\bibinfo{year}{2021}).

\bibitem{irfan2025quantum}
\bibinfo{author}{Irfan, A.~K.} \& \bibinfo{author}{Yu, C.}
\newblock \bibinfo{journal}{\bibinfo{title}{Quantum approximate optimization
  algorithm: Performance on simulators and quantum hardware}}.
\newblock {\emph{\JournalTitle{arXiv preprint arXiv:2509.24213}}}
  (\bibinfo{year}{2025}).

\bibitem{erdds1959random}
\bibinfo{author}{Erd\H{o}s, P.} \& \bibinfo{author}{R\'enyi, A.}
\newblock \bibinfo{journal}{\bibinfo{title}{On random graphs i}}.
\newblock {\emph{\JournalTitle{Publ. math. debrecen}}}
  \textbf{\bibinfo{volume}{6}}, \bibinfo{pages}{18} (\bibinfo{year}{1959}).

\bibitem{chen2023transmon}
\bibinfo{author}{Chen, L.} \emph{et~al.}
\newblock \bibinfo{journal}{\bibinfo{title}{Transmon qubit readout fidelity at
  the threshold for quantum error correction without a quantum-limited
  amplifier}}.
\newblock {\emph{\JournalTitle{npj Quantum Information}}}
  \textbf{\bibinfo{volume}{9}}, \bibinfo{pages}{26} (\bibinfo{year}{2023}).

\bibitem{chi2026variational}
\bibinfo{author}{Chi, Y.} \emph{et~al.}
\newblock \bibinfo{journal}{\bibinfo{title}{Variational quantum algorithms with
  invariant probabilistic error cancellation on noisy quantum processors}}.
\newblock {\emph{\JournalTitle{Science China Physics, Mechanics \& Astronomy}}}
  \textbf{\bibinfo{volume}{69}}, \bibinfo{pages}{210312}
  (\bibinfo{year}{2026}).

\bibitem{wang2025depth}
\bibinfo{author}{Wang, Z.}, \bibinfo{author}{Mandell, J.}, \bibinfo{author}{Xu,
  Y.} \& \bibinfo{author}{Shi, J.}
\newblock \bibinfo{journal}{\bibinfo{title}{A depth-independent linear chain
  ansatz for large-scale quantum approximate optimization}}.
\newblock {\emph{\JournalTitle{arXiv preprint arXiv:2509.17296}}}
  (\bibinfo{year}{2025}).

\bibitem{maciejewski2025improving}
\bibinfo{author}{Maciejewski, F.~B.}, \bibinfo{author}{Biamonte, J.},
  \bibinfo{author}{Hadfield, S.} \& \bibinfo{author}{Venturelli, D.}
\newblock \bibinfo{journal}{\bibinfo{title}{Improving quantum approximate
  optimization by noise-directed adaptive remapping}}.
\newblock {\emph{\JournalTitle{Quantum}}} \textbf{\bibinfo{volume}{9}},
  \bibinfo{pages}{1906} (\bibinfo{year}{2025}).

\bibitem{natanael2026exhaustive}
\bibinfo{author}{Natanael, N.}, \bibinfo{author}{Sugiarto, H.} \&
  \bibinfo{author}{Tjandra, Y.}
\newblock \bibinfo{journal}{\bibinfo{title}{Exhaustive entanglement-topology
  enumeration and edge-vertex-level attribution in quantum feature map design
  for classification}}.
\newblock {\emph{\JournalTitle{Journal of Computing Theories and
  Applications}}} \textbf{\bibinfo{volume}{4}}, \bibinfo{pages}{454--474}
  (\bibinfo{year}{2026}).

\bibitem{kim2023cuda}
\bibinfo{author}{Kim, J.-S.} \emph{et~al.}
\newblock \bibinfo{title}{Cuda quantum: The platform for integrated
  quantum-classical computing}.
\newblock In \emph{\bibinfo{booktitle}{2023 60th ACM/IEEE Design Automation
  Conference (DAC)}}, \bibinfo{pages}{1--4} (\bibinfo{organization}{IEEE},
  \bibinfo{year}{2023}).

\end{thebibliography}



\section*{Acknowledgements}
GPU-enhanced computations were performed in Calvin Institute of Technology, Jakarta. Y.T. and H.S.S. were supported by PT Lancs Arche Consumma (MOU.003/CIT/XI/2022). A.B.S. was funded by Hibah P2MI ITB (Project ID : STEI.PPMI-1-30-2026).

\section*{Author contributions statement}

Y.T. and H.S.S. conceived the idea and designed the experimental setup. A.B.S. developed the theoretical framework and validated the experimental setup. C.V.S. and J.G.T. implemented the simulated experiments and collect the data. Y.T. designed, implemented, and interpreted the noise threshold modeling. H.S.S and A.B.S validated the noise threshold modeling. Y.T. performed the computation on real quantum hardware. H.S.S and A.B.S contributed to the interpretation of the results. Y.T. wrote the first draft of the manuscript. All authors contributed to the revised manuscript and approved the final version.

\section*{Data Availability statement}
All datasets and source codes encompassing all phases of this study: the computational experiments, the threshold extraction modeling, and the figures/diagram generation; will be made publicly available on GitHub (\url{https://github.com/QCRG-CRC/QAOA-Effective-Noise-Threshold}) upon the manuscript acceptance.




\end{document}